\documentclass{mn2e}
\usepackage{times}

\input{psfig.sty}

\newif\ifAMStwofonts

\title[Beyond the standard model: coupled magnetic disc--corona solutions]
        {Beyond the standard accretion disc model: coupled magnetic disc--corona
solutions with a physically motivated viscosity law}
\author[A. Merloni]
        { A. Merloni
\\Max-Planck-Institut f\"ur Astrophysik,
        Karl-Schwarzschild-Strasse 1, D-85741, Garching, Germany 
}

\date{}

\begin{document}

\maketitle

\label{firstpage}

\begin{abstract}
We present a systematic, analytical study of geometrically thin, 
optically thick accretion disc solutions
for magnetized turbulent flows, with an 
$\alpha$-like viscosity prescription. 
Under the only assumptions that (1) Magneto-Rotational 
instability (MRI) generates the turbulence that produces the
anomalous viscosity needed for accretion to proceed, and that (2) the
magnetic field amplified by the instability saturates due to buoyant
vertical escape, we are able to self-consistently solve the disc
structure equations including the fraction of power $f$ that is
carried off by
vertical Poynting flux (and likely dissipated {\it outside} the optically
thick disc). For low-viscosity discs, we obtain stable high-$f$ solutions at
low accretion rates, when gas pressure dominates, and unstable,
low-$f$, radiation pressure dominated solutions at high
accretion rates. For high viscosity discs, instead, a new {\it
  thermally and viscously stable},
radiation pressure dominated solution is found, characterized by
$f\sim 1$ and appearing only above a critical accretion rate (of
the order of few tenths of the Eddington one). We discuss the regimes of
validity of our assumptions, and the astrophysical relevance of our solutions.
We conclude that our newly discovered thin disc solutions, 
possibly accompanied by powerful, magnetically
dominated coronae and outflows, should be seriously considered as
models for black holes accreting at super-Eddington rates. 
\end{abstract}


\section{Introduction}
The so-called standard model for accretion discs \cite{ss73} describes
an optically thick, geometrically thin solution to the accretion
equations in which all our ignorance about the mechanisms of angular
momentum transport and turbulent viscosity is side-stepped by making
the (physically reasonable) {\it Ansatz} that the turbulent stress
tensor scales with the local pressure, with a constant of
proportionality that we will denote here $\alpha_{\rm SS}$. 
Such an approach proved enormously productive, and despite the great
development of numerical models in the last decade, standard disc
modeling still remains at the very heart of accretion disc
phenomenology and the basic link between theory and observations.

The recent evolution of numerical studies, instead,  has proved itself
fundamental to shed light on the very nature of the $\alpha_{\rm SS}$
prescription, by elucidating the crucial role of MHD turbulence for
the enhanced transport properties of accretion discs \cite{bh98}. In
particular, non-radiative flows are ideal objects of study by means of 
global three dimensional simulations (see Balbus \& Hawley 2002 and
reference therein).

On the other hand, the observed spectrum of thin accretion discs around many
compact objects suggests  that a corona is also present \cite{np94}. Such
coronae are most likely a by-product of the internal disc dynamics (as
in the model originally proposed by  Galeev, Rosner \& Vaiana 1979)
and local 3-D MHD simulations of stratified discs \cite{ms00},
although adopting a thermodynamics which is most likely inappropriate
for the problem at hand, may support this idea.

The main aim of the present work is to incorporate, in the simplest
possible way, our current knowledge on the properties of turbulent MHD
flows in the standard accretion disc theory.
This will allow us, on the one hand, to extend the predictive power of
the standard model by
self-consistently including in the theory the magnetic properties of
the flow and the generation of a magnetic corona by means of vertical
Poynting flux; and on the other
hand, to consider MHD turbulent discs in a regime where global, 3-D 
numerical simulations  are
currently unable to make predictions, i.e. for highly luminous,
radiatively efficient thin discs. 

Thus, we assume the accretion proceeds through a geometrically thin disc,
{\it \`a la} Shakura \& Sunyaev (1973), and we employ vertically
integrated equations. 
Furthermore, we assume that magneto-rotational instability (MRI; see
Balbus \& Hawley 1998, and reference therein) is 
the primary source of the turbulent viscosity and, consequently, that the
turbulent magnetic stresses responsible for the angular momentum
transport in the disc scale with magnetic pressure.
Then, the accretion disc structure is fully described
once the relationship between magnetic and disc pressure
(given either by gas or radiation) is established.

The issue
of the viscous scaling can be translated
into the uncertainty about the mechanisms by, and the level at, which the
disc magnetic field saturates. In the following, we make the
working assumption (in a sense justified {\it a priori} by the need of
explaining the observed presence of strong coronae in many accretion
powered systems) that magnetic field is disposed of mainly by
vertical buoyancy \cite{sr84,sc89}, and verify a posteriori its validity. 
Then we will show how under this condition, and taking into account 
the property that MRI
grows less rapidly in radiation dominated discs, 
the scaling of magnetic pressure can be determined. This in turn
specifies the closure relation that fully
describes a coupled accretion disc--corona system. 

Furthermore, we will show that 
by solving the structure equations 
a new solution is found, which has no analog in the standard
Shakura-Sunyaev theory, and is relevant for highly viscous (and highly
magnetized) flows accreting at a rate of the order of (or above) the
Eddington one.

\section{Basic equations}
Numerical simulations of MHD turbulent discs clearly show that the
non-linear outcome of the MRI is a fully three dimensional,
anisotropic turbulence, that exhibits strongly correlated fluctuations
in the azimuthal and radial components of velocity (Reynolds stresses)
and magnetic field (Maxwell stress). In accordance with those results,
and for the sake of simplicity, we will assume that the
vertically averaged anomalous stress is dominated by Maxwell stresses,
and therefore
\begin{equation}
\tau_{R \phi} \simeq k_0 P_{\rm mag},
\end{equation}
where $P_{\rm mag}=\frac{B^2}{8\pi}$ is the magnetic pressure, 
$k_0$ is a factor of order unity that depends on the relative intensity
of radial and azimuthal field components\footnote{It is worth stressing that 
such an expression for the anomalous stress can be recovered exactly
from the full MHD equations of a turbulent (thin) disc after a careful
averaging procedure is carried on \cite{bgh94,bp99,bla02}.}.

This in turn implies that the local heating rate in a thin Keplerian  
disc is given by
\begin{equation}
\label{eq_heat}
Q_+=\frac{3}{2}c_{\rm s} \tau_{R \phi} \simeq \frac{3}{2} k_0 c_{\rm s}P_{\rm mag},
\end{equation}
where $\Omega_{\rm K}$ Keplerian angular velocity, $c_{\rm s}=\sqrt{P_{\rm tot}/\rho}$ is the isothermal sound
speed. 

The growth rate of MRI is influenced by the ratio
of the gas to magnetic pressure, as demonstrated both by analytical
studies of the linear regime of the instability \cite{bs01} and by
numerical MHD simulations of radiation pressure dominated discs
\cite{tss02}: both found that, due to the substantial compressibility
of MHD turbulence in radiation pressure dominated discs, 
\begin{equation}
\label{eq_bs01}
\sigma \simeq \Omega_{\rm K} \frac{c_{\rm g}}{v_{\rm A}}, 
\end{equation}
where $c_{\rm g}=\sqrt{P_{\rm gas}/\rho}$ is the gas sound speed. 
The magnetic field,
with initial (highly subthermal) amplitude $B_0$, rapidly grows to $B \sim
B_0 e^{\sigma t_{\rm b}}$, where $H$ is the total disc scaleheight and
\begin{equation}
\label{eq_tb}
t_{\rm b}\simeq H/2v_{\rm D}
\end{equation} 
is  the mean buoyant rise-time. In order to estimate the
saturation field, an estimate for the upward drift velocity $v_{\rm
  D}$ caused by buoyancy of magnetic field is needed. In the case of
uniform discs permeated by a net mean field, 
where the magnetic field can be modeled as a collection
of flux tubes that retain their individuality over time scales much
longer than dynamical time, it has been shown \cite{sr84} that
$v_{\rm D} \simeq v_{\rm A} (c_{\rm g}/c_{\rm s}) (\pi/2 C_{\rm
  D})^{1/2} (a/H)^{1/2}$. Here  $v_{\rm A}= |{\bf B}| (4 \pi
\rho)^{-1/2}$ is the Alfv\'en speed, $a$ is a flux tube cross section
and $C_{\rm D}$ a drag coefficient, whose numerical value can be
estimated to be $\sim 1$--$10$. If the tube cross section $a$ were
independent on the ratio of gas to radiation pressure, such an
expression for the upward drift velocity would lead to a final scaling
of magnetic pressure with the gas pressure only, as already argued in
Stella \& Rosner (1984). However, the applicability of a flux tube
analysis to the case of MRI turbulent flows is not justified.
The work of Stone et al. (1996) clearly demonstrates that magnetic
concentrations, although always present, do not persist as coherent
structures.
Moreover, as already mentioned, 
the turbulence in radiation pressure dominated discs is
compressible. Then, in the vertical fluid momentum equation 
any vertical gradient of the 
magnetic pressure will not be balanced by either radiation (because
radiation is diffusive) nor gas
pressure (because azimuthal field energy density may exceed the true
gas pressure, see below). Vertical motions are excited, with
approximately (see e.g. Eq. 4.19 in Blaes 2002)
\begin{equation}
\rho\frac{v_{\rm D}}{t_{\rm b}} \sim  \frac{B_{\phi}^2}{4 \pi H},
\end{equation}
which would imply (for a mainly azimuthal field) a direct
proportionality between the vertical drift velocity and the Alfv\`en
speed.
We therefore assume
\begin{equation}
\label{eq_vdrift}
v_{\rm D} = b v_{\rm A},
\end{equation}
with $b$ constant of the order of unity.

If we then put together Eqs. (\ref{eq_bs01}), (\ref{eq_tb}) and
(\ref{eq_vdrift}),  we have that 
$P_{\rm mag} \ln(B/B_0) \propto \sqrt{P_{\rm tot}P_{\rm gas}}$, where 
$P_{\rm tot}=P_{\rm gas}+P_{\rm rad}$ is the sum of gas plus
radiation pressure.
Neglecting the logarithmic dependence on the initial field, 
this in turn suggests that, in MRI dominated turbulent flows, the field will saturate at 
an amplitude such that \cite{tl84,bur85,szu90,mf02}:
\begin{equation}
\label{eq_visc}
P_{\rm mag}=\alpha_0 \sqrt{P_{\rm gas}P_{\rm tot}},
\end{equation}
where we have introduced the constant
\begin{equation}
\label{eq_alpha}
\alpha_0=\frac{1}{\beta}\left(\frac{P_{\rm tot}}{P_{\rm gas}}\right)^{1/2},
\end{equation} 
where $\beta=P_{\rm tot}/P_{\rm mag}$ is the usual plasma parameter.

It is worth noting here the relationship between the constant $\alpha_0$ 
and the Shakura-Sunyaev coefficient $\alpha_{\rm SS}\equiv P_{\rm
  mag}/P_{\rm tot}$ (which is {\it not} constant in our theory).
We have
\begin{equation}
\label{eq_a_ss}
\alpha_0=\alpha_{\rm SS} \sqrt{\frac{P_{\rm tot}}{P_{\rm gas}}}.
\end{equation}
The condition that the turbulence be subsonic ($v_{\rm A} \la c_{\rm
  s}$) implies $\alpha_{\rm
  SS} \la 1/2$ (or, equivalently, $\beta \ga 2$),  
and from (\ref{eq_a_ss}) we see that  
$\alpha_0$ can still be larger than unity, 
when radiation pressure dominates in the disc.

The magnetic flux escaping in the vertical direction may dissipate a
substantial fraction of the gravitational binding energy of the
accreting gas {\it outside}  the optically thick disc, with obvious
deep implications for the spectrum of the emerging radiation \cite{hm91}.
How much of this flux really escapes into the low-density environment
above and below the disc (in the so-called {\it corona}), depends on
the details of vertical transport of magnetic flux tubes and vertical
stratification of the disc, which is ignored in our model. 
Nonetheless, we can safely assume that the fraction  
of power dissipated in the corona
is determined by the ratio $f$ of the vertical Poynting flux
 to the local heating rate $Q_+$
\cite{sz94}.
The simplest way to estimate the vertical Poynting flux is to just
assume that  
\begin{equation}
F_{\rm P}\simeq v_{\rm D} P_{\rm mag}.
\end{equation}
This translates into (see Eq. \ref{eq_heat})
\begin{equation}
\label{eq_f}
f=\frac{v_{\rm A}}{k_1 c_{\rm s}}=\sqrt{\frac{2}{k_1^2 \beta}}
\simeq \sqrt{2 \alpha_0/k_1} \left(1+\frac{P_{\rm
      rad}(R)}{P_{\rm gas}(R)}\right)^{-1/4}.
\end{equation}

In the following, we will assume that the numerical factor 
(of the order of unity) $k_1=3k_0/2b=1$. 
Differences in its exact value will affect the
numerical value of the critical viscosity parameter $\alpha_0$ (see
below), but won't affect the nature of the solutions we find. 

Thanks to the closure relation (\ref{eq_f}), we are now 
able to build a self-consistent accretion disc--corona solution. In
order to do so, exactly as in the case of the standard Shakura-Sunyaev
solution, we have to solve simultaneously four 
conservation equations (conservation of vertical momentum, mass,
angular momentum and energy) and supplement them with 
the equations of state for the pressure and opacity as functions of 
density and temperature. The only modification with respect to the
standard derivation is the reduction  of the locally
dissipated energy by the factor $(1-f)$ \cite{sz94}, 
with $f$ given by Eq.~(\ref{eq_f}), and the adoption of the viscosity
law (\ref{eq_visc}).

For simplicity, we present here the solutions in Newtonian
approximation, as in Shakura \& Sunyaev (1973), with an efficiency
$\epsilon_0=0.082$ appropriate for non-rotating black holes (a full general relativistic analog of the equations
below can be found in the appendix of Merloni \& Fabian 2002b).
We concentrate only on the innermost regions of the disc, where
opacity is dominated by electron scattering ($\kappa_{\rm es} = 0.4$
cm$^2$ g$^{-1}$).
Introducing dimensionless units for the central mass ($m \equiv M/M_{\odot}$), 
accretion rate ($\dot m\equiv \epsilon_0 \dot M c^2/L_{\rm Edd}$) 
and distance from the center ($r \equiv R/R_{\rm S}=Rc^2/2GM$),
we obtain, for the radiation pressure dominated solution, the
following expressions for disc scaleheight, pressure (in dyne), density  
(in g cm $^{-3}$), both calculated at the disc mid-plane, 
and temperature (in K):

\begin{eqnarray}
\label{eq_rad}
h &=&\frac{H}{R_{\rm S}} \simeq 11 \dot m J(r)  (1-f) \nonumber \\
P & \simeq & 4.6 \times 10^{18}  (\alpha_0 m)^{-8/9}  r^{-8/3} 
(\dot m J(r))^{8/9}  (1-f)^{4/5} \nonumber \\
 \rho  & \simeq & 1.9 \times 10^{-4}  (\alpha_0 m)^{-8/9}  r^{3/9} 
(\dot m J(r))^{-10/9}  (1-f)^{-2}  \nonumber \\
 T & \simeq & 2.1 \times 10^8   (\alpha_0 m)^{-2/9}  r^{-2/3} 
(\dot m J(r))^{2/9}  
\end{eqnarray}
supplemented with the closure equation for $f$:
\begin{equation}
\label{eq_fr}
 \frac{4 \alpha_0^2 -f^4}{f^4(1-f)^2}= 7.3
\times 10^5 (\alpha_0 m)^{2/9}  r^{-7/3} 
(\dot m J(r))^{16/9}.
\end{equation}
 
Analogously, in the gas pressure dominated part of the disc, we have:
\begin{eqnarray}
\label{eq_gas}
h&\simeq& 2.3 \times 10^{-2} (\alpha_0 m)^{-1/10}
r^{21/20} (\dot m J(r))^{1/5}  (1-f)^{1/10}  \nonumber \\
 P & \simeq & 2.7 \times 10^{18} (\alpha_0 m)^{-9/10}
 r^{-51/20} (\dot m J(r))^{4/5} (1-f)^{-1/10}  \nonumber \\
 \rho &  \simeq &  19  (\alpha_0 m)^{-7/10}  r^{-33/20} (\dot
m J(r))^{2/5}  (1-f)^{-3/10} \nonumber \\
 T & \simeq & 8.0 \times 10^8  (\alpha_0 m)^{-1/5}  r^{-9/10} 
(\dot m J(r))^{2/5}  (1-f)^{1/5} 
\end{eqnarray}
together with
\begin{equation}
\label{eq_fg}
 \frac{4 \alpha_0^2 -f^4}{f^4(1-f)^{9/10}} = 4.7 \times 10^2
(\alpha_0 m)^{1/10}  r^{-21/20} (\dot m J(r))^{4/5}.
\end{equation}

In the above formulae, the function
$J(r)=1-\sqrt{r_{\rm in}/r}$, 
with $r_{\rm in}=3$ corresponding to the innermost stable circular
orbit of a free particle around a non-rotating black hole, describe
the no-torque at the inner boundary condition.

\section{Study of the solutions}

To find all the family of possible solutions, we proceed in the
following way:
for each set of the parameters ($m$, $\dot m$, $r$ and $\alpha_0$) we
find the roots of both Eq.~(\ref{eq_fr}) and (\ref{eq_fg}) in the
interval $0<f<1$. We then
substitute the obtained values of $f$ into the appropriate 
expressions for the gas
and radiation pressure and finally discard all the non-consistent
solution (i.e. those found with the formula for gas pressure dominated
discs that gives $P_{\rm rad}>P_{\rm gas}$ and vice versa).

The value of the viscosity parameter $\alpha_0=k_1/2=0.5$ is a
critical point for the set of equations describing the coupled
disc--corona system, in that it separates two qualitatively different 
regimes. In the next two sections we will examine these two regimes,
beginning with the low viscosity case.

\subsection{$\alpha_0 < 1/2$: an almost standard solution}

\begin{figure}
\psfig{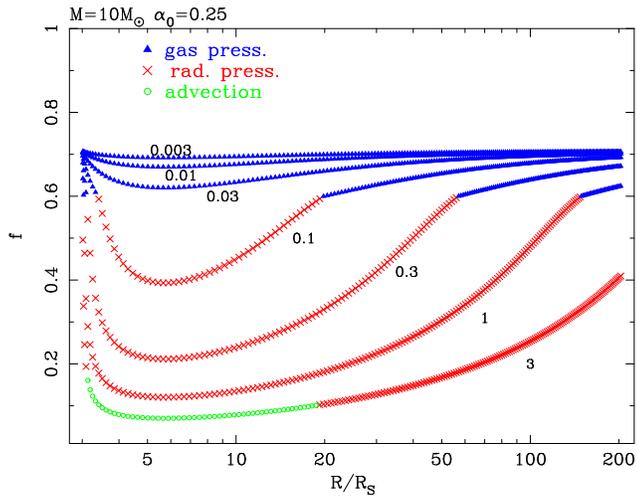}
\caption{Radial profiles of the fraction of
power transported vertically by Poynting flux, $f$ for $m=10$ and
$\alpha_0=0.25$ and different values of the accretion rate: $\dot m=
0.003,0.01,0.03,0.1,0.3,1,3$. The disc structure has been computed
using Newtonian approximation for non-rotating black hole. Filled
triangles represent gas pressure dominated solutions; crosses
radiation pressure dominated ones, while empty circles mark the
solutions where cooling by advection should be taken into account
(slim disc solutions).} 
\label{fig_lowalpha}
\end{figure}

It turns out that for $\alpha_0 < 1/2$ each set of parameters define a
unique solution. Its properties have already been discussed
\cite{mf02}, but we present them here in a more general framework.
The fraction of disc power dissipated in the corona, $f$, 
tends to its maximum value, $\sqrt{2\alpha_0}$,
when gas pressure dominates (low accretion rates), 
and decreases as the accretion rate
increases and radiation pressure
becomes more and more important. 
This is shown in Fig.~\ref{fig_lowalpha}, where, for $m=10$ and
$\alpha_0=0.25$, we plot the radial dependence of the coronal fraction
for different values of the accretion rate. 

As in the standard case, radiation pressure dominated part of
the disc are thermally and viscously unstable, although the combined
effects of the modified viscosity law (\ref{eq_visc}) and the
stabilizing effect of the corona, reduce the extent of the unstable
region of the parameter space \cite{szu90}. 

\subsection{$\alpha_0 > 1/2$: a new stable solution at high accretion rates}

The nature of the solutions changes qualitatively if $\alpha_0> 1/2$.
Because in gas pressure dominated parts of the disc $f\sim
\sqrt{2\alpha_0}$, there can be no solutions there with $f<1$. That is
to say that the condition $\alpha_0<1/2$ should  hold in gas pressure
dominated discs, and these high viscosity solutions can only exist if
radiation pressure dominates. 

\begin{figure}
\psfig{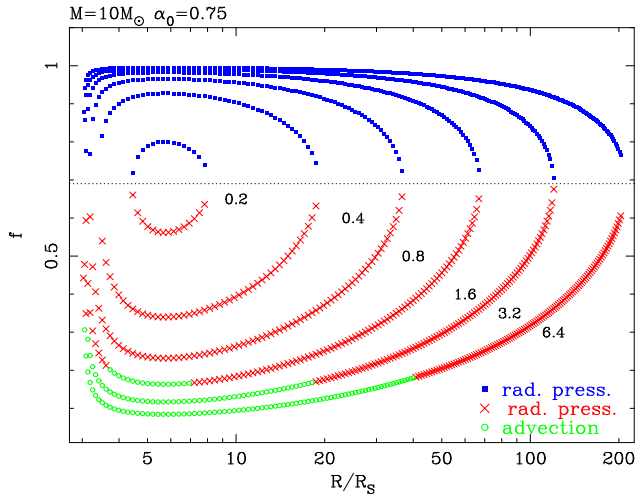}
\caption{Radial profiles of the fraction of
power transported vertically by Poynting flux, $f$ for $m=10$ and
$\alpha_0=0.75$ and different values of the accretion rate: $\dot m=
0.2,0.4,0.8,1.6,3.2,6.4$. The disc structure has been computed
using Newtonian approximation for non-rotating black hole. Filled
squares represent stable, radiation pressure dominated solutions; crosses
unstable ones, while empty circles mark the
solutions where cooling by advection should be taken into account
(slim disc solutions). The dotted horizontal line mark the critical
value $f_{\rm cr}\simeq 0.689$ that separates the two solutions.} 
\label{fig_highalpha}
\end{figure}

For each value of
$\alpha_0 > 1/2$ the accretion rate $\dot m(r,f)$ (Eq.~\ref{eq_fr}) has a minimum for
$f \in (0,1)$. Thus, solutions exists only for accretion rates larger
that a critical value, $\dot m_{\rm cr}(r)= \min_{f \in (0,1)}[\dot m
(f,r)]\equiv \dot m (r,f_{\rm cr})$. This corresponds to a critical
value, $f_{\rm cr}$, of the fraction of power vertically transported by
Poynting flux. For $\dot m(r) > \dot m_{\rm cr}(r)$
two radiation pressure dominated 
solutions are found for each set of parameters,
one with $f<f_{\rm cr}$ and the other with $f>f_{\rm cr}$ 
(see Fig.~\ref{fig_highalpha}). The
first is a ``standard'' (unstable) one, 
characterized by a low value of the fraction of power transported
vertically by Poynting flux, and is just
the continuation of the radiation pressure dominated branch discussed
in the previous section at high
values of $\alpha_0$. The second one, which is discussed here for the
first time,
appears due to the feedback effect of the closure relation (\ref{eq_f}).
It is a (mildly) radiation pressure dominated solution endowed with a
powerful corona, in that 
$f\rightarrow 1$ as the accretion rate increases. 

The values of $\dot m_{\rm cr}(r)$ and $f_{\rm cr}$ can be found by
solving
$\partial \dot m(r,f) / \partial f =0$, or, equivalently, the equation
 $f^5-12\alpha_0^2 f+8 \alpha_0^2 =0$. For every $r$, re obtain $f_{\rm cr}
\simeq 0.7417$ for $\alpha_0=0.5$, quickly converging towards $f_{\rm
  cr}=2/3$ for increasing $\alpha_0$. The critical accretion rate,
though, depends on $r$. The lowest possible value of the critical
accretion rate, that can be simply denoted $\dot m_{\rm cr}$, 
appears at $r\simeq 4.64$ (which is the radius where
the function $J(r)^{16/9}r^{-7/3}$ has a minimum).
We have numerically computed $\dot m_{\rm cr}$ as a function of
$\alpha_0 $. The relation can be approximated by the analytic
expression
\begin{equation}
\dot m_{\rm cr}=0.34 \alpha_0 m^{-1/8},
\end{equation}
which is accurate to better than 3\% for $\alpha_0 > 1$ and to better than
10\% for $\alpha_0 > 0.6$. The above formula also makes manifest the
main differences that would be found if we chose supermassive black
holes instead that stellar mass ones. Namely, the critical accretion
rate separating gas from radiation pressure dominated solutions
and 
that determining the existence of our new high viscosity, high-$f$ solution 
would be reduced by a factor $m^{-1/8}$. The qualitative properties of
the solutions would be left unchanged.

Fig.~\ref{fig_highalpha} shows, for $m=10$ and
$\alpha_0 = 0.75$, the radial dependence of the coronal fraction
for different values of the accretion rate. 
The two solutions appear for $\dot m \ga 0.19$ and show an opposite trend
as the accretion rate increases: while in the first case $f$ decreases
and the disc becomes less dense and more radiation pressure dominated,
in the second case $f$ increases, the disc becomes denser and less
radiation pressure dominated.

It is worth studying the stability property of this newly discovered
solution. Thermal stability requires \cite{pir78}
\begin{equation}
\left(\frac{\partial \ln Q_+}{\partial \ln
    T}\right)_{\tau}<\left(\frac{\partial \ln Q_-}{\partial \ln
    T}\right)_{\tau}.
\end{equation}
We have $(\partial \ln Q_-/\partial \ln T) = 4$ and, from
(\ref{eq_rad}), 
$Q_+ \propto \dot m (1-f) \propto T^{9/2} (1-f)$. Therefore, the
stability criterion becomes:
\begin{eqnarray}
\label{eq_sta}
\left(\frac{\partial \ln Q_+}{\partial \ln
    T}\right)_{\tau}&-&\left(\frac{\partial \ln Q_-}{\partial \ln
    T}\right)_{\tau}= \frac{9}{2}\left(1+\frac{\partial (1-f)}{\partial
    \dot m}\right)-4 \nonumber \\ &=& \frac{1}{2}+\frac{4f
    (4\alpha_0^2-f^4)}{(f^5-12 f \alpha_0^2+8\alpha_0^2)}<0.
\end{eqnarray}
Because the numerator of the second term is always positive for
$\alpha_0>1/2$, we have that for $f>f_{\rm cr}$ (which is a pole for
the second term in the above equation) indeed the criterion is
satisfied, while the opposite is true for $f<f_{\rm cr}$.
On the other hand, viscous stability requires that 
$(\partial \dot M/\partial \Sigma)>0$. We have $\Sigma \propto \rho H
\propto \dot m^{-1/9} (1-f)^{-1}$, and the stability condition can be
written as
\begin{equation}
\label{eq_sta_v}
\frac{2}{9}\left(\frac{1}{2}+\frac{4f
    (4\alpha_0^2-f^4)}{f^5-12 f \alpha_0^2+8\alpha_0^2}\right)<0,
\end{equation} 
which is satisfied when thermal stability condition is also satisfied.

The  very important fact that our new discovered high-$f$ solutions
are both viscously and thermally stable may be an indication that,
whenever allowed to do so, the system will chose those solutions over
the low-$f$ ones.
The real physical state of a high viscosity disc at high accretion
rates, though, is likely more complicated and inherently time
dependent.
Indeed, the very fact that for $\alpha_0>0.5$ no gas pressure
dominated solutions exist, make us believe that in the outer part of
the disc magnetic turbulence will self-regulate to keep the viscosity
parameter below its critical value. Thus, 
the value of the viscosity parameter may not be constant throughout the
disc, but more likely increases inwards, as also suggested by
high-resolution numerical MHD simulations of the inner parts of 
non radiative discs \cite{ra01,kh02}. 
For any given radial profile of the viscosity parameter $\alpha_0$, 
the radius at which it reaches its critical value
will therefore play an important role in determining the physical
properties of the disc. We may speculate that in a full time dependent
solution  both the
viscosity parameter and the local accretion rate will adjust in order
to avoid discontinuities in the flow there, but the exact nature of
the final disc structure is impossible to predict from our stationary 
solutions.

\section{Discussion}
Beside the thin disc approximation, which is straightforward to verify
{\it a posteriori} for all our solutions but the low-$f$, high-$\dot
m$ ones (as in the standard SS case), our results are rooted in two
further assumptions. The first is that the growth rate of MRI is given
by (\ref{eq_bs01})  in radiation pressure dominated discs,
the second is that buoyancy is the main mechanism of field saturation
in turbulent magnetized discs. Let us examine these two issues one at
a time.

As discussed in Blaes \& Socrates (2001), 
the applicability of Eq.~(\ref{eq_bs01}) depends on the capability of
radiative diffusion to destroy radiation sound waves on the length
scale of the MRI, and requires that the azimuthal magnetic field
component dominates the vertical one, as should be the case in 
thin discs with Keplerian velocity profiles 
(see also the numerical simulations of initially random fields in
Hawley, Gammie \& Balbus 1996).
Let us define the Alfv\'en speed associated to the vertical field
component, $v_{{\rm A}z}=B_z (4 \pi \rho)^{-1/2}$, and the (small)
parameter $\delta\equiv v_{{\rm A}z}/v_{\rm A}$. The diffusion
wavenumber for Keplerian discs is \cite{bs01}:
\begin{equation}
\label{eq_kdiff}
k_{\rm diff} \equiv \left(\frac{3 \Omega_{\rm K} \kappa_{\rm es} \rho}{c}\right)^{1/2}.
\end{equation} 

Then, Eq.~(\ref{eq_bs01}) should be valid if the characteristic
wavenumber of the instability, $\Omega_{\rm K}/v_{{\rm A}z}$ is of the
order of (or larger than) the diffusion wavenumber $k_{\rm diff}$.
By using  the expression (\ref{eq_rad}) for the density, (\ref{eq_f}),
and the definition of $v_{{\rm A}z}$, we obtain the following
condition for the validity of our assumption:

\begin{equation}
k_{\rm diff} f H \delta \simeq 65 f
m^{1/18} \left(\frac{\dot m J(r)}{\alpha_0}\right)^{4/9} r^{-21/36} 
\delta \la 1.
\end{equation}   
Noting that $\max[J(r)^{4/9} r^{-21/36}] \simeq 0.2$, this
translates into\footnote{We adopt here
  the notation $A=10^x A_x$.}
\begin{equation}
\label{cond_diff}
1.5 f m_1^{1/18} \left(\frac{\dot m}{\alpha_0}\right)^{4/9} \delta_{-1} \la 1,
\end{equation}
which is always satisfied for the low-$f$ radiation pressure dominated
solutions, while, for the $f\sim 1$ solutions, it breaks down for very
high accretion rates 
$\dot m \gg 0.55 \alpha_0 m^{-1/8} \delta_{-1}^{-9/4} \simeq 1.6 \dot
m_{\rm cr} \delta_{-1}^{-9/4}$. It is clear that in order for our
viscosity scaling (\ref{eq_visc}) to be applicable to the
high-$f$ solutions at super-Eddington rates, higher azimuthal field
strengths (and low $\delta$) are required. 
However, if  $\dot m$ grows too large, and Eq.~(\ref{cond_diff}) is
not satisfied, then the
magnetic stresses, and the magnetic pressure, will be proportional to
the total pressure. The condition of subsonic turbulence would then imply 
again $\alpha_0<1/2$. The fraction of power transported vertically
by Poynting flux, $f$ will tend again to its maximum value $\sim
\sqrt{2\alpha_0}$, as in the gas-pressure dominated
solutions. Therefore, it seems plausible to envisage that the high-$f$
solutions could extend to arbitrarily large accretion rates,
regardless to the viscosity law and the detailed properties of the
MRI field, as long as $\alpha_0$ is sufficiently high in these regimes.      

The second main assumption that was used in the derivation of our
solutions was to consider buoyancy the main mechanism by which the
disc gets rid of the magnetic field. In fact, two other competing
mechanisms should be considered: turbulent diffusion and dissipation.  

Let us first compare the typical buoyancy timescale
($t_b\simeq H/2 b v_{\rm A}$) 
with the turbulent diffusion one
($t_{\rm d}\simeq H^2/\nu$). The kinematic viscosity coefficient
$\nu$ can be recovered from the standard viscous form of the stress to
obtain $\nu=2\tau_{R\phi}/(3 \rho \Omega_{\rm K})$. We then have 
\begin{equation}
\frac{t_{\rm b}}{t_{\rm d}} = \frac{k_0 v_{\rm A}}{6 b c_{\rm s}}= \frac{k_1^2
  f}{6},
\end{equation} 
Therefore, for any physical self-consistent, $f<1$ solution, 
buoyancy is always a more efficient mechanism of field escape than
field diffusion, unless $k_1=3 k_0/2b$ is substantially larger than unity..

On the other hand, the field may dissipate by
cascading down the subsonic turbulence and leave its energy inside the
disc.
The typical timescale for this process is set by the coherence time of
the turbulent flow, that we have assumed to be $\tau_{\rm t} \equiv
1/\sigma \sim
v_{\rm A}/(c_{\rm g}\Omega_{\rm K})$. This would give 
\begin{equation}
\frac{t_{\rm b}}{\tau_{\rm
    t}}=\frac{1}{4 b \alpha_0},
\end{equation}
which implies that our working assumption is justified only for large
viscosity solutions. In practice, if $\alpha_0$ is small, the actual
maximal fraction of power dissipated into the corona will be further reduced
by a factor $1+0.25/b \alpha_0$. However, the high viscosity, high-$f$
solutions discussed in section 3.2 will be little affected by
dissipational losses of energy inside the disc. 


\section{Conclusions}

We have presented a full analysis of thin accretion disc solutions
under the assumptions that 
the magnetic field generated and
amplified by MRI (which is at the origin of MHD turbulent stress in
the disc) saturates mainly due to buoyant transport in the
vertical direction. From these assumptions, a viscosity prescription is
derived, in which magnetic pressure, and therefore
turbulent stresses, scale proportionally to the geometric
mean of gas and total (gas plus radiation) pressure. This prescription,
in turn, allows us to uniquely determine, for every set of the
parameters ($m$, $\dot m$, $r$ and $\alpha_0$) the fraction of the accretion
power $f$ which is transported vertically  by Poynting flux. 
For low values of the viscosity parameter the  solution is unique,
with $f$ tending to its maximum value,
$\sqrt{2\alpha_0}$,
when gas pressure dominates (low accretion rates), 
and decreasing as the accretion rate
increases and radiation pressure
becomes more and more important. 
For $2\alpha_0>1$ there are no solutions for gas pressure
dominated regions, while in the radiation dominated part of the flow
two solutions are possible for every value of the accretion rate. The
first is a ``standard'' (unstable) one, with $f \ll 1$, while the second
has $f\rightarrow 1$ (corona dominated) 
as the accretion rate increases and is thermally and
viscously stable. 


 
Two main results from the present study are important for the interpretation of
observed spectral energy distributions 
of accretion-powered systems. The first is the property
of low-$\alpha_0$ solutions to be more corona-dominated (i.e. to have
the largest possible value of $f$) at low accretion rates. This has been
discussed in Merloni \& Fabian (2002a) in the context of low-luminosity
black hole. It is indeed an observational fact that the relative
strength of the hard, optically thin, power-law-like component in
their X-ray spectra increases with respect to the optically thick,
quasi-thermal one as $\dot m$ decreases.
The second is the existence of a new high-$f$ solution at high
accretion rates for high values of the viscosity parameter. 

As in the
case of corona-dominated solutions at low $\dot m$, it is plausible to
speculate that the large Poynting fluxes in the vertical direction
that characterize these solutions are to be associated with the
generation of powerful coronae and, possibly, powerful outflows/jets,
too \cite{mf02}. 
Indeed,
there are observational hints that black holes believed to accrete at
Eddington or super-Eddington rates bear many similarities with their
low-rate counterparts.  This class of objects may include galactic
black holes in the {\it Very High} state
(see Done 2002, and references therein), whose X-ray spectra
show a power-law component, likely of non-thermal origin \cite{pc98,gie99}, 
often as strong as the quasi-thermal one;
 broad line radio galaxies, where the weakness
of X-ray reflection features has been interpreted as evidence for highly
ionized discs surrounded by powerful coronae at high accretion rates \cite{brf02};
powerful blazars, where it has been shown \cite{gc02} how
the relativistic jet in these sources may dominate the total output
power at very high (inferred) $\dot m$. Furthermore, some  Ultra Luminous X-ray
sources (ULX; Makishima et al. 2000) have shown power-law like spectra,
analogous to those of black hole candidates in their very high state,
at inferred accretion rates of a few times the Eddington rate for a
30$M_{\odot}$ black hole \cite{kdm02}.

In order to assess the relevance of our new solution in all the
above cases, vertically integrated models should
be abandoned, and
the complicated problem of determining the disc structure in the vertical
direction, together with the disc--corona
boundary, tackled. 
This would allow us to establish the relationships between the
(vertically integrated) parameter $f$, 
and the observationally determined fraction of power dissipated in the
corona.
In particular, it is only by carefully studying the disc--corona
transition region that will be possible to assess how super-Eddington
(if at all) our new high $\alpha_0$ solutions can be. In principle, if
a high fraction of accretion power could be dissipated in the corona
without disrupting the optically thick disc,
there would be no limitation to the total luminosity that the source
could emit, because the radiation is shining on the disc from outside. 
Such a possibility to obtain super-Eddington luminosities is
physically different from the one recently proposed by Begelman (2001;
2002), although both depend on the existence of magnetized radiation
pressure dominated thin discs. In fact, the non-linear analysis of 
Gammie (1998) and Blaes \& Socrates' (2001) photon bubble instability
carried on in Begelman (2001) is applicable to the low-$f$ radiation
dominated solutions we have presented here, which is much more strongly
radiation pressure dominated than the high-$f$ ones, while the thinner
discs of the high-$f$ solutions have almost equal radiation and gas
pressures, and therefore should not be too strongly inhomogenous. 

Finally, we point out that the existence of multiple solutions for the
same set of parameters (high $\alpha_0$ case), may give rise to
interesting hysteresis effects during spectral transition caused by
long term variation of the accretion rate \cite{mc03}.
A full time-dependent
analysis of our solution is needed in order to make definite
predictions in this sense.

\section*{Acknowledgments}
I thank Tiziana Di Matteo, Andy Fabian and Rashid Sunyaev for their
valuable suggestions and the anonymous referee for carefully reading
the manuscript and for his useful comments.

\bsp

\label{lastpage}


\begin{thebibliography}{}



\bibitem[Balbus, Gammie \& Hawley 1994]{bgh94}
Balbus, S. A., Gammie, C. F. \& Hawley, J. F., 1994, MNRAS, 271, 197.

\bibitem[Balbus \& Hawley 1998]{bh98}
Balbus, S. A. \& Hawley, J. F., 1998, Rev. Mod. Phys. 70, 1.

\bibitem[Balbus \& Hawley 2002]{bh02}
Balbus, S. A. \& Hawley, J. F., 2002, to appear in ``Turbulence and
Magnetic Fields in Astrophysics, eds. E.Falgarone \& T.Passot. astro-ph/0203353

\bibitem[Balbus \& Papaloizou 1999]{bp99}
Balbus, S. A. \& Papaloizou, J. C. B., 1999, ApJ, 521, 650.

\bibitem[Ballantyne, Ross \& Fabian 2002]{brf02}
Ballantyne, D. R., Ross, R. R. \& Fabian, A. C., 2002, MNRAS, 332, L45 

\bibitem[Begelman 2001]{beg01}
Begelman, M. C., 2001, ApJ, 551, 897 

\bibitem[Begelman 2002]{beg02}
Begelman, M. C., 2002, ApJL, 568, L97

\bibitem[Blaes 2002]{bla02}
Blaes, O., 2002, to appear in "Accretion Disks, Jets, and High Energy
Phenomena in Astrophysics", Proceedings of Session LXXVIII of Les
Houches Summer School, Chamonix, France, eds. F.Menard, G. Pelletier,
G.Henri, V. Beskin, and J. Dalibard (EDP Science: Paris and Springer:
Berlin). astro-ph/0211368

\bibitem[Blaes \& Socrates 2001]{bs01}
Blaes, O. \& Socrates, A., 2001, ApJ, 553, 987.


\bibitem[Burm 1985]{bur85}
Burm, H., 1985, A\&A, 143, 389.


\bibitem[Done 2002]{don02}
Done, C., 2002, Phil. Trans. of the Royal Society A360, 1967.
astro-ph/0203246.

\bibitem[Galeev, Rosner \& Vaiana 1979]{grv79}
Galeev, A. A., Rosner, R. \& Vaiana, G. S., 1979, ApJ, 229, 318

\bibitem[Ghisellini \& Celotti 2002]{gc02}
Ghisellini, G. \& Celotti, A., 2002, in ``Issues in Unification of
AGN'', eds. 
R. Maiolino, A. Marconi, N. Nagar, ASP Conf. Proc., 258, 273
                 

\bibitem[Gierli\'nski et al. 1999]{gie99}
Gierli\'nski, M., Zdziarski, A. A., Poutanen, J., Coppi, P.,
 Ebisawa, K. \& Johnson, W. N., 1999, MNRAS, 309, 496.

\bibitem[Haardt \& Maraschi 1991]{hm91}
Haardt, F. \& Maraschi, L., 1991, ApJL, 380, L51.



\bibitem[Hawley, Gammie \& Balbus 1996]{hbg96}
Hawley, J. F., Gammie, C. F. \& Balbus, S. A., 1996, ApJ, 464, 690.



\bibitem[Krolik \& Hawley 2002]{kh02}
Krolik, J. H. \& Hawley, J. F., 2002, ApJ, 573, 754

\bibitem[Kubota, Done \& Makishima 2002]{kdm02}
Kubota, A., Done, C. \& Makishima, K., 2002, accepted for publication
in MNRAS. astro-ph/0209113

\bibitem[Maccarone \& Coppi 2003]{mc03}
Maccarone, T. J. \& Coppi, P., 2002, MNRAS, 338,189



\bibitem[Makishima et al. 2000]{mak00}
Makishima, K. et al, 2000, ApJ, 535, 632 

\bibitem[Merloni \& Fabian 2002a]{mf02}
Merloni, A. \& Fabian, A. C., 2002, MNRAS, 332, 165

\bibitem[Merloni \& Fabian 2002b]{mf02b}
Merloni, A. \& Fabian, A. C., 2002b, MNRAS, submitted.

\bibitem[Miller \& Stome 2000]{ms00}
Miller, K. A. \& Stone, J. M., 2000, ApJ, 534, 398

\bibitem[Nandra \& Pounds 1994]{np94}
Nandra, K. \& Pounds, K. A., 1994, MNRAS, 268, 405



\bibitem[Piran 1978]{pir78}
Piran, T., 1978, ApJ, 221, 652.

\bibitem[Poutanen \& Coppi 1998]{pc98}
Poutanen, J. \& Coppi, P. S., 1998, Physica Scripta, T77, 57.

\bibitem[Reynolds \& Armitage 2001]{ra01}
Reynolds, C. S. \& Armitage, P. J., 2001, ApJL, 561, L81.


\bibitem[Sakimoto \& Coroniti 1989]{sc89}
Sakimoto, P. J. \& Coroniti, F. V., 1989, ApJ, 342, 49.

\bibitem[\protect\citename{Shakura \& Sunyaev }1973]{ss73}
Shakura, N. I. \& Sunyaev, R.A., 1973, A\&A, 24, 337.


\bibitem[Stella \& Rosner 1984]{sr84}
Stella, L. \& Rosner, R.,  1984, ApJ, 277, 312.

\bibitem[Stone et al. 1996]{sto96}
Stone, J. M., Hawley, J. F., Gammie, C. F. \& Balbus, S. A., 1996,
ApJ, 463, 656.

\bibitem[\protect\citename{Svensson \& Zdziarski }1994]{sz94}
Svensson, R. \& Zdziarski, A. A., 1994, ApJ, 436, 599.

\bibitem[\protect\citename{Szuszkiewicz }1990]{szu90}
Szuszkiewicz, E., 1990, MNRAS, 244, 377

\bibitem[\protect\citename{Taam \& Lin }1984]{tl84}
Taam, R. E. \& Lin, D. N. C., 1984, ApJ, 287, 761

\bibitem[Turner, Stone \& Sano 2002]{tss02}
Turner, N. J., Stone, J. M. \& Sano, T.,  2002, ApJ, 566, 148.

\end{thebibliography}
\end{document}